\documentclass[twoside,a4paper]{article}
\usepackage{fullpage, comment}
\usepackage{amsfonts,amssymb,amsmath,amsthm}
\usepackage{tikz}
\usetikzlibrary{shapes}
\theoremstyle{definition}
\usepackage{bussproofs}
\usepackage{listings}
\usepackage{makecell}
\usepackage{multicol}
\usepackage{caption, hyperref}
\newcommand{\toto}{xxx}

\newcommand{\vir}[1]{``#1''} 
\usepackage{xargs}                      
\usepackage{xcolor}  
\usepackage[colorinlistoftodos,prependcaption,textsize=tiny]{todonotes}

\newcommandx{\YAG}[2][1=]{\todo[linecolor=black,backgroundcolor=teal!25,bordercolor=teal,#1]{\textcolor{teal}{\textbf{Yackolley}}\\#2}}
\newcommandx{\ST}[2][1=]{\todo[linecolor=black,backgroundcolor=red!25,bordercolor=red,#1]{\textcolor{teal}{\textbf{Sara}}\\#2}}
\newcommandx{\AD}[2][1=]{\todo[linecolor=blue,backgroundcolor=blue!25,bordercolor=blue,#1]{\textcolor{teal}{\textbf{Antonella}}\\#2}}
\newcommandx{\MP}[2][1=]{\todo[linecolor=green,backgroundcolor=green!25,bordercolor=green,#1]{\textcolor{teal}{\textbf{Zay}}\\#2}}

\usepackage[normalem]{ulem}
\newcommand\redout{\bgroup\markoverwith
               {\textcolor{red}{\rule[0.5ex]{2pt}{0.8pt}}}\ULon}

\makeatletter
\let\latex@@line\line
\def\line{\@ifnextchar(\latex@@line{\hbox to\hsize}}
\makeatother

\usepackage{amsthm}

\def \trans #1#2#3#4{ #1~{\xrightarrow{(#3,~#2)}}~#4 }
\def \iss #1#2 {\mbox{\small #1:=}#2}
\def \siss #1#2#3{\mbox {\small $#1\nearrow#2$:=}#3}

\def\attest#1#2{#1 \vartriangleright #2}

\title{Pluralize: a Trustworthy Framework for High-Level Smart Contract - Draft}
\author{Zaynah Dargaye \and Antonella Del Pozzo \and Sara Tucci-Piergiovanni \\ \\ CEA, LIST, PC 174, Gif-sur-Yvette, 91191, France \\ \{lea-zaynah.dargaye,antonella.delpozzo,sara.tucci\}@cea.fr}

\begin{document}
 
	\newcounter{linecounter}
	\newcommand{\linenumbering}{(\arabic{linecounter})}
	\renewcommand{\line}[1]{\refstepcounter{linecounter}
		\label{#1}
		\linenumbering}
	\newcommand{\resetline}{\setcounter{linecounter}{0}}
	
	\maketitle
	\begin{abstract}
The paper presents \textit{Pluralize} a formal logical framework able to extend the execution of  blockchain transactions to events coming from external oracles, like  external time,  sensor data, human-made declarations, etc. These events are by essence \textit{non-reliable}, since transaction execution can be triggered by information whose veracity cannot be established by the blockchain.  To overcome this problem, the language features a first-order logic and an authority algebra to allow formal reasoning and establish \textit{accountability } of agents  for blockchain-enabled transactions. 
We provide an accountability model that allows to formally prove the accountability of agents by a formal proof  locally executable by each agent of the blockchain. 
\end{abstract}


	\section{Introduction} \label{sec:intro}

Blockchains promise to increase the level of trust among actors that
do not trust each other. This promise is often misunderstood,
especially when smart contracts come into play. Smart contracts are
pieces of code that run in an actively replicated fashion in the
blockchain, but current platforms do not offer any native support for
software verification/validation, so that bugs can affect the smart
contracts behavior. Moreover, in many industrial applications smart
contracts react to information coming from external sources (sensors,
human interfaces, other applications), often called $oracles$ in the
blockchain jargon, which amplifies their vulnerability.

Another difficulty is that smart contracts are commonly associated to
Ethereum-based ones, which depart considerably from the transactional
nature of protocols they should support. Those contracts propose
indeed a diametrical opposite paradigm where the smart contract plays
the role of mediator between parties. Very popular are, for instance,
many forms of auctions, where a smart contract $auction$ will be
called by any wallet -- a client -- to register a bet and the smart
contract will gather all the bets and elect the winner. In a way the
concept of an application-level \vir{transaction} between symmetric
parties, triggered only when certain conditions are met, must be
twisted in a very different semantics because a transaction becomes a
client call to the smart contract. The mismatch between paradigms
create a lot of confusion and difficulties to use smart contracts in a
proper way. 

We advocate that the first step for better design, and hence safety
and security, of smart contracts is to stick as much as possible to
the Bitcoin transactional model. In Bitcoin, transactions are
transfers of ownership of tokens -- $Bitcoins$ -- through both
asymmetric cryptography and a replicated shared memory -- the {\it
blockhain} --, which stores enough information in order to
recompute/validate each transfer. This memory is shared by agents part
of a peer-to-peer network, where each agent is represented by an
address, i.e. the wallet.  A transfer of a quantity of Bitcoins from a
wallet to another is guarded by a condition that has to be satisfied
-- {\textit{a guard}}. For the Bitcoin system to be safe, guards must
be identically evaluated by each peer. Currently, identical and
correct evaluation of guards relies on the {\it autonomous} evaluation
of the guard by each peer, which is deterministic and closed to the
current knowledge of the blockchain\footnote{Note that such guards are
encoded in Bitcoin scripts, called Opcodes that compile to an abstract
machine that every peer has locally.}. More specifically, guards in
Bitcoin can trigger the execution of a transaction upon the evaluation
of simple conditions, as signature integrity and timelocks, stating
that a transfer must be authorized by the owner of the transferred
tokens and after a certain time (in general measured as a number of
blocks). 

\subsection{Motivation}
{
 In~\cite{Herlihy18}, the author presents a protocol that involves different parties in a transfer of a car and crypto-currencies ownership in a sort of chained transactions loop such that either all the transactions take places or none.
 Let us consider the following scenario:
 \begin{itemize}
 	\item Carol wants to sell a Cadillac for bitcoin;
 	\item Alice wants to buy Carol's Cadillac, but with alt-coin;
 	\item Bob wants to trade alt-coin for bitcoin.
 \end{itemize} 


  All the parties do not trust each other and the protocols explicit how to enforce trust between them thanks to cryptography. The motivation example in ~\cite{Herlihy18} specifies the transfer of an asset via smart contracts: scripts published on the blockchain that establish and enforce conditions necessary to transfer the asset from an agent $A$ to another agent $B$. In particular $A$ publishes a smart contract that locks the asset to transfer along with two conditions to execute the transfer to $B$: a hashlock $h$ and a timelock $t$. Hashlock $h$ means that if $B$ provides a secret $s$ such that $H(s)=h$ before $t$ expires then the asset ownership is transferred from $A$ to $B$. If the timelock $t$ is not satisfied, the asset returns to $A$. The timelock enables to encode synchronization and sequentialition of the smart contracts transactions. 
 So, Alice will transfer her alt-coin to Bob that will convert them into bitcoin and transfer them to Carol that will transfer her ownership title for the Cadillac to Alice.

 The trade takes place on the blockchain, there is no third trusted-party, and Alice,Bob and Carol do not trust each other. Hence, they use smart contract to secure and enforce the transactions between them. {Let $\Delta$ be the time for one agent to publish a smart contract on a blockchain and for the other agent to detect the change. For simplicity, in this work we consider that all parties have perfectly synchronized clocks.} The protocol becomes:
 \begin{itemize}
   \item Alice generates a secret $s$, and publishes the transaction of her alt-coin to Bob
 triggered by the hashcodelock of $s$, $h=H(s)$, and a timelock of $6\Delta$;
   \item When Bob confirms that Alice's smart contract has been published, Bob publishes
 a contract that transfer to Carol the bitcoin trigerred by $h$ and the timelock of $5\Delta$;
  \item When Carol confirms that Bob's smart contract has been published, Carol publishes a smart contract that transfer the Cadillac ownership title to Alice with the same $h$ and a timelock of $4\Delta$;
  \item When Alice confirms that Carol's smart contract has been published, Alice send the secret $s$ to unlock the $h$ and trigger the transaction of the Cadillac ownership title;
  \item Carol sends $s$ to Bob in order to obtain her bitcoin;
  \item Bob sends $s$ to Alice in order to obtain his alt-coin.
 \end{itemize}

%

Thanks to the refund in case of timelock expiration, this protocol works even if one agent halts. Timelock are used to enforce the sequential order of transactions. Finally, the only irrational behavior that can corrupt the protocol is Alice revealing the secret too earlier and in that case she is the unique victim of it. This protocol is an implementation of smart contracts that are autonomously executes on blockchains and that establishes and enforces contractual conditions. However, nothing bad happend as long as we consider only digital assets, i.e., if something goes wrong, then the digital asset ownership goes back to the previous owner. 
Now, let us consider the following scenario in which Alice instead of buying the Cadillac is renting it with the condition that is in a good state. In this case it is necessary to have an attestation declaring the state of the car as an extra condition such that the transaction take place in the blockchain. Suppose that the Cadillac is equipped with IoT sensors so that the information about the state is automatically computed and provided to the blockchain as a transaction. Those sensors work like a specific agent in the protocol. Such computation, being out of the blockchain is not reliable. It follows that the current state of the Cadillac may not be good when Alice receives it. Then, Alice as well, can provide her observation of the current state of the Cadillac. Then the protocol has to enable Alice to take countermeasures if needed, e.g., being refund and return the Cadillac. 

In such context, a naive protocol can be the following:
\begin{itemize}
 \item Alice generates a secret $s$, and publishes the transaction of her alt-coins to Bob triggered by the hashcodelock of $s$, $h$, and a timelock of $7\Delta$, and a declaration about the good state of the Cadillac from its IoT sensors;
 
\item When Bob confirms that Alice's contract has been published, Bob publishes a contract that transfer to Carol the bitcoin trigerred by $h$, the timelock of $6\Delta$;

\item When Carol confirms that Bob's smart contract has been published, Carol publishes a smart contract that state the renting contract between her and Alice with the same $h$, a timelock of $5\Delta$ and a declaration about the good state of the Cadillac from its IoT sensors;

\item When the IoT sensors confirm that Carol's smart contract has been published, the IoT sensors compute and publish a smart contract declaring the Cadillac current state with the same  $h$ and a timelock of $4\Delta$; 

\item When Alice confirms that Carol's and IoT sensors smart contracts have been published, Alice observes the Cadillac. Two cases can occur:
\begin{itemize}
	 \item the IoT sensors declaration matched with the Alice observation, then she accepts the renting contract and sends the secret $s$ to unlock the $h$;
	 \item Carol sends $s$ to Bob in order to obtain her bitcoin; 
	 \item Bob sends $s$ to Alice in order to obtain his alt-coin.
\end{itemize}
Otherwise:
\begin{itemize} 	 
	\item otherwise Alice does not release the secret $s$.
\end{itemize}
%
%
 
\end{itemize}

This naive protocol presents some limitation. First of all, when we consider the physical world we have to estimate in a different manner the timelock. Alice might take more than $\Delta$ time to receive and observe the Cadillac. Second, in such scenario there are two unreliable conditions that trigger transactions: IoT sensors computation and Alice observation, both performed outside of the blockchain. Thus, it is not possible to establish who is telling the truth: Alice may lie and/or the IoT sensors can be bugged. Naively, assuming declarations out of the blockchain as true as trigger of transactions may introduce contradiction in the blockchain, leading to inconsistencies and lost of reliability of the information in the blockchain.
IoT sensors, as the Alice observations, are referred as Oracle in the blockchain jargon. Triggers (transactions) coming from physical world Oracles cannot be validated against all the blockchain information and can not be evaluated as valid by default. 
To avoid having contradictions in the blockchain the validation of such triggers has guarantees that adding a new information into the blockchain is not going to invalidate something that is already present in the blockchain. To do so, at each validation of a claim $claim$, the validation mechanism of the blockchain $\mathbb{V}$ has to make sure that no already validated transaction is invalidated by $claim$. In that case, $\mathbb{V}$ returns the list of discord -- the list of claims that are in contradiction. We call this computation the {\it proof-of-discord}. In our example, the proof-of-discord returns IoT sensors declaration and Alice observation. The {\it proof-of-discord} enables to manage conflicts. 
Notice, the conflict management, being application and physical world dependent, is out of the scope of this paper. 

}

\subsection{Our Contribution}
The main challenge for smart contracts is to extend the
Bitcoin paradigm to guards whose evaluation that are not closed to the
blockchain, going far beyond the simple semantics of Bitcoin scripts
and matching requirements of most industrial applications.
This possibility rises the problem of the \textit{reliability} of the
guards execution, that can be triggered upon \textit{lies} declared by
the external agent/oracle, not detectable at the time of validation by
blockchain peers.

To solve this issue, we introduce a logical framework
called \textit{Pluralize}.  Pluralize features a formal language,
$Plurality$, for smart contracts expressed as token transfers with
guards whose evaluation is not closed to the blockchain. To cope with
the possible unsafe execution of those non-closed
guards, \textit{Plurality} semantics relies on a \textit{formal
accountability model for transactions}, based on
Cyberlogic~\cite{RS:HCSS03} for the management of agents
accountability. More specifically, we establish an accountability
model to infer accountability of agents basing on their actions:
submitting, validating, executing transactions, providing additional
information for non-closed guards.  In case of logical contradictions,i.e. discords, 
thanks to the Cyberlogic authority algebra and our Coq theorem prover,
it is possible to trace back the {\it chain of trust}, i.e., formally
proving the agents accountable for the discording logical 
properties through a formal mathematical certificate. Being a
certificate means that the formal proof is locally checkable by any
agent, for this reason we will call it in the reminder of the
paper \textit{proof-of-discord}.

The main technical challenge in this \textit{demarche} is the
formalization of the accountability model because it depends on the
execution model of transactions, which in turn depends on the
execution and validation blockchain model. Since the $Pluralize$
framework wants to be technology agnostic, i.e. be instanciable on
different types of blockchain, it is not possible to built upon a
specific execution and validation model, as Bitcoin or Ethereum.

To overcome that difficulty we propose an operational semantics of the
$Plurality$ language built upon the blockchain abstraction provided
in\cite{DBLP:journals/corr/abs-1802-09877}, called the Blockchain
Abstract Data Type (ADT). Following
the \cite{DBLP:journals/corr/abs-1802-09877}'s construction, each
submitter $A$ of the transaction $T: A \rightarrow B$ that wants to
append the transaction to the blockchain has to successfully call the
blockchain ADT. Interestingly, the blockchain ADT allows to pass as
parameter application specific validation rules for transactions,
i.e. the guards, and blockchain specific rules, such as the longest
chain rule, for blockchain consistency.

The paper is organised as follows: Section~\ref{sec:over} gives an overview of Pluralize, Section~\ref{sec:back} introduces
the technical background, Section~\ref{sec:mf} the language to define
transaction-based protocols along with its semantics based on
Cyberlogic and the Blockchain ADT, Section~\ref{sec:relatedwork}
discusses related work and Section~\ref{sec:conclusion} concludes the
paper.

       \section{Overview} \label{sec:over}
{
Pluralize is a formal framework that provides cooperation and composition of formal methods in order to conciliate the gap between higher-level smart contracts – expectations of the collective belief -- and smart contracts in the blockchain ecosystem. Pluralize features a dedicated formal language, {\it Plurality}, that is more suitable to higher-level smart contracts. Plurality features claimed and closed guards. Pluralize enriches a host blockchain in order to benefit its information reliability with more complex validation mechanisms. The validation mechanism is enriched with a first-order-logic (FOL) interpreter for closed guards and proof-of-discord for claimed guards. To reach the same level of reliability that current bitcoin-like smart contract, (i) Plurality smart contract are certified-by-construction, (ii) proof-of-discord are also certified and (iii) the execution model of Pluralize is based on a abstraction compatible with the distributed nature of the blockchain. 

\paragraph{Validation of Closed Guards}
Currently, the autonomously evaluated guards of bitcoin-like transaction are scripts.  A script is a sequence of opcodes that is able to verify on cryptographical conditions.  Enriching the script expressiveness with first-order boolean logic closed to the blockchain knowledge increases the autonomy and reliability of smart contract while decreasing the need of oracles. Pluralize features closed guards.  At implementation level, A Plurality smart contract  comes with its set of logical definitions, the set of predicates needed to validate its closed guards. To evaluate validity of closed guards, Pluralize enriches the validation mechanism $\mathbb{V}$ of the host blockchain with an interpreter for first-order logic. This interpreter knows the set of logical definitions of the smart contract 
and evaluates closed guards. 

\paragraph{Validation of Claimed Guards}
Reliability of information coming from agents that works in \vir{black-box} style involved in a shared computation is addressed in security thanks to accountability and trust management. Accountability designates the responsibility of an agent for an information. In case of problem, it is then possible to identify the agents involved by tracing back the chain of trust: computing the list of elementary data involved in the problem and identify the agents that are accountable for them. A claimed guard is very similar to a computation in "black-box" style: it is a statement of an agent and its validity is guaranteed by this agent: its accountable.

A trust management tool is a policy language -- a formal language dedicated to formalize security policy of a whole system -- that enables to address several security language in a same policy language. A Trust Management (introduces inPolicyMaker~\cite{Blaze98}), features an authority algebra that enables to explicit the accountability of the different agents or components involved in a system.

Pluralize features claimed guards. At implementation level, Plurality is a dialect of a trust management logical framework, {\it Cyberlogic}~\cite{RS:HCSS03} presented in~\ref{sec:back}. A claimed guard is a claim $claim$ of a property $p$ by an agent $A$, denoted $\attest{A}{p}$ in Plurality (as in Cyberlogic). A claim is validated only if it is not in discord with an already validated claims. $\Gamma_{claim}$ denotes the already validated claims.  To validate a claim $claim$ Pluralize enriches $\mathbb{V}$ with the proof-of-discord mechanism. If $claim$ is in discord with $\Gamma_{claim}$, $\mathbb{V}$ returns the list of claims that are in discord.  the validation of $claim$ is its confrontation with $\Gamma_{claim}$. If $\Gamma_{claim}$ satisfies $\neg claim$ then $claim$ is in discord with $\Gamma_{claim}$.

As Cyberlogic, Plurality features a first-order logic that enables to reason and drive proof on the authority algebra. $\Gamma_{claim} \vdash \neg claim$ means $\Gamma_{claim}$ satisfies $\neg claim$: there is a constructive proof in Plurality that satisfies $\neg claim$.  The set of claims $\gamma_{claim}$ that are in discord with $claim$ is the subset of $\Gamma_{claim}$ used in the proof of $\Gamma_{claim} \vdash \neg claim$. If no proof of  $\Gamma_{claim} \vdash \neg claim$ is realised then $claim$ is validate.

\paragraph{Trustworthy of The Validation Mechanism}
Pluralize enriches the validation mechanism $\mathbb{V}$ of an host blockchain $\mathbb{B}$. Pluralize enrichment has to be reliable and to preserve the reliability  of $\mathbb{B}$. Hence Pluralize execution model has to be able to explicit both (i) $\mathbb{V}$ and (ii) the enrichment. A blockchain is a shared data structure that ensures the immutability of its contains. A blockchain ideally ensures that every agents shared the history (the same contains). In reality, it is now well known that for some blockchain, as bitcoin, several histories might exist: it is possible to have forks. In a context where a sequence of transaction is driven by external component, as in the Cadillac example, the existence of several histories can be handle out of the chain. It is then possible to define an execution model as a unique history. Pluralize features claimed and closed guards in order to leverage the autonomy of smart contracts execution through several transactions. In that case, it is not possible to make same assumption. The possibility to have several histories is a property of the host blockchain $\mathbb{B}$. Furthermore, Plurality is not just a DSL built upon smart contracts of $\mathbb{B}$, it requires the enrichment of its validation mechanism $\mathbb{V}$ which is that is defined at $\mathbb{B}$ protocol level. Hence, Pluralize execution model has to be able to abstract underlying properties of $\mathbb{B}$ regarding its capacity of forking and explicit $\mathbb{V}$. Pluralize does not mange the non unicity of the blockchain as it inherits it from $\mathbb{B}$. However, Pluralize cannot make the assumption that its execution model relies on an ideal blockchain.

 Abstract Data Structure~\cite{perrin:hal-01286755}, is a formal
 specification tool that enables to formalize shared data structure in
 order to address both (i) the execution model of the data structure
 for interaction with the agents and (ii) the data structure owned
 properties as a distributed protocol.

 Using an execution model from such a formal specification improves the trustworthy of certified-by-design smart contracts.  \cite{DBLP:journals/corr/abs-1802-09877} presents such a formal framework for blockchain. In particular,~\cite{DBLP:journals/corr/abs-1802-09877} explicits the abstraction of the validation mechanism of the blockchain that is suitable to formalize the enrichment that {\it Pluralize} proposed. Furthermore, such abstraction enables to deal with concurrent histories -- viewing a blockchain as a tree -- in the execution model. That is complete different from the tradition abstraction of blockchain as an execution model: a sequence of blocks -- one unique history. The latter formalization is known as too strong for bitcoin protocol.

 Pluralize execution model refers to the blockchain ADT~\cite{DBLP:journals/corr/abs-1802-09877}. 
 A quick description is provided in~\ref{sec:back}. In Figure \ref{fig:pluralize} is depicted the overview described so far. From a transactional protocol are obtained (i) the Plurality transactional protocol, with closed and claimed guards and (ii) the logical definition for the closed guards. From those we derive the validation mechanism $\mathbb{V}$ enriched with a FOL interpreter for closed guards (e.g., agent $A$ transfers organic cotton to agent $B$, $B$ transfers organic dress to agent $C$.) and proof-of-discord for claimed guards (e.g., IoT sensors agent asses that the Cadillac is in a good state). Proof-of-discord fetches from the blockchain the claims already validated. Finally, $\mathbb{V}$ publishes on the blockchain only validated information. In case of information invalidation by proof-of-discord, $\mathbb{V}$ returns to the conflict manager (out of the scope of this work) the list of discording claims.

\begin{figure}
	\begin{tikzpicture}
		
		\node[] () at (0.75,2.3) {Transactional protocol};
		\draw(0,0) rectangle (1.5,2); 
			\draw (0,1.5) -- (1.0,1.5);
			\draw (0,.3) -- (1.0,.3);
			\draw (0,.6) -- (.5,.6);
			\draw (0,.9) -- (1.3,.9);
			\draw (0,1.2) -- (1.3,1.2);
			
		\node[draw] (triggers) at (8,1.5)[align=left] {Logical definition for \\ the closed guards};
		
		\draw(1,-3) rectangle node[align=left] {Plurality transactional protocol \\ with closed and claimed guards} (6,-1) ;
		
		\node [] () at (10.5,-3.7) {$\mathbb{V}$};
		\draw(5.5,-4) rectangle (11.5,-6);
		
		\draw(5.7,-4.2) rectangle node[align=left] {FOL \\ Interpretator} (8.4,-5.8);
		\draw(8.6,-4.2) rectangle node[align=left] {Proof-of-discord} (11.3,-5.8);
		
		\draw(11,-3) [dotted]rectangle node{Conflict manager} (14, -1);
		
		\draw(0,-7) rectangle node{Blockchain data structure} (7,-8);
		
		\draw[->] (2,1.3) -- (6,1.3);
		\draw[->] (1.2,-.2) -- (1.2,-.9);
		\draw[->] (8.2,.9) -- (8.2,-4.5);
		\draw[->] (6.2,-3.1) -- (7.5, -3.9);
		\draw[->] (7, -6.1) -- (4,-6.9);
		\draw[->] (5.5, -6.9)--(9.5, -5.7);
		
		\draw[dotted,->] (11,-3.9)--(12,-3.2);
		
 	\end{tikzpicture}
 	\caption{Pluralize framework. The dotted components are outside the scope of this work. }\label{fig:pluralize}
\end{figure}
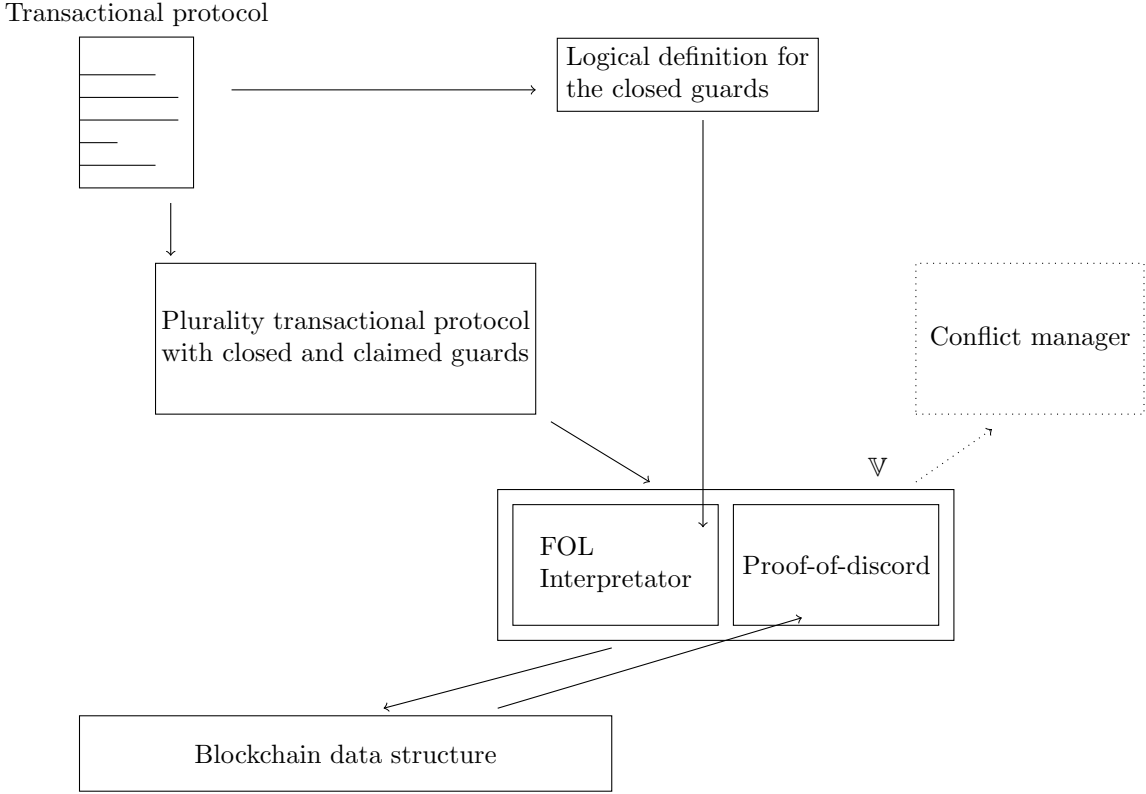

}

	\section{Technical Background} \label{sec:back}
		{ 

 The {\it Pluralize} framework is a formal framework for
 accountability of blockchain transactions based on
 Cyberlogic~\cite{RS:HCSS03}. {\it Pluralize} consists of (i) a host
 blockchain $\mathbb{B}$ (the blockchain target for the smart contract
 execution), (ii) execution and validation mechanisms for transactions
 --enriching $\mathbb{B}$ with guards based on accountability-- and
 (iii) a formal language, {\it Plurality}. This language allows to
 define smart contracts featuring such guards. {\it Plurality} is a
 shallow-embedding in the Coq theorem prover, being executable and
 providing a proof environment for establishing
 proof-of-discord.  {\it Pluralize} features as well an
 execution model based on the Blockchain ADT. In the following we
 present the technical background of {\it Pluralize}, introducing more
 in detail Cyberlogic along with the principles to establish a
 proof-of-discord in Section \ref{sec:tb:fam} and the
 Blockchain ADT in Section \ref{ssec:adt}.

\subsection{Formal Accountability Management}\label{sec:tb:fam}

\paragraph{Cyberlogic.} 
    Cyberlogic~\cite{RS:HCSS03} is a trust management formal
    framework. Traditional trust management tools enables to
    specify and monitor security property.  Cyberlogic is a formal
    language that features both an authority algebra and a first-order
    logic, which allows to specify and verify distributed protocols
    with explicit accountability of agents,
    called \textit{authorities}.

    Cyberlogic formula: $\attest{A}{d}$, also called a $claim$,
     states that $A$ claims $d$, a property, and therefore $A$
    is accountable for it. 
    Cyberlogic is particularly adequate to specify
    transactional protocols endowing cryptography mechanisms where
    transactions react to non-reliable events~\cite{RS:HCSS03}. In
    particular, in the context of a transaction where the satisfaction
    of the guard $gd$ triggers the transaction execution, it is
    possible to model a satisfied guard $gd$ as a claim; the validity
    of the guard is then endorsed by an agent:
    $\attest{\Omega_k}{gd}$ means that $\Omega_k$ claims that $gd$ is
    true. 

\cite{cyber18} presents a shallow-embedding of Cyberlogic in the Coq
theorem prover that makes it possible to certify Cyberlogic
protocols and claims. 
The Coq theorem prover realizes the isomorphism of Curry-Howard --
{\it typing is proving}: it is based on Calculus of Inductive
Construction CiC~\cite{Cic} \footnote{CiC is $\lambda-$calculus with
types as first class citizens}. In such theorem prover
proofs are programs that are computable and checkable on any
machine. Such proof are formal certificate. 
{\it Plurality} is a dialect of Cyberlogic, means that every smart contract
design and implemented in Plurality are
certified-by-design. Furthermore, the claimed guards are also
Cyberlogic claims, means that every reasonning on the authority
algebra is also certified.

\paragraph{The proof-of-discord.}
  The logical rules of Cyberlogic enable to reason on the
  authority algebra in a manner that it is possible to trace back the
  chain of trust, leveraging on the fact that the derivation tree
  contains Cyberlogic claims. Technically, it is to trace-back in
  the derivation tree of the claims used for establishing a property. In
  particular, a property $p$ claimed by an authority $A$ that does not
  contradict any hypothesis of the proof context is considered as an
  axiom. The reduction of the proof context to the hypotheses,
  involved in the proof, is the tracing-back of the chain of trust. The
  Cyberlogic implementation in Coq enables to formally certify the
  chain of trust of Cyberlogic protocols or claims. Tracing back the
  chain of trust enables to formally detect conflicts and point-out
  the accountable agents in the conflicts. A conflict is a set of
  Cyberlogic claims that are in contradiction with each other. The
  agents involved in this set are the potential accountable for this
  contradiction. Detecting that a property $p$ is in contradiction
  with a set of claims $\Gamma$ means to verify that $\Gamma$
  satisfies $\neg p$ denoted by $\Gamma \vdash \neg p$.  If
  the verification succeeds by a proof $\pi$, then the claims of $\pi$
  are in contradiction with $p$.
  It is exactly how the proof-of-discord work with the validation of claimed guards.

\paragraph{Accountability at run-time.} \label{ssec:cps}
If on one hand Pluralize defines a formal language for transaction
protocols based on blockchains, on the other hand the
proof-of-discord needs to store specific information in the
blockchain and mechanisms to deal with agent actions at run-time.  In
particular, claimed guards must be continually checked at run-time
against the blockchain state. Logical contradictions must be detected
automatically, but the proof-of-discord can be obtained only
with an interactive theorem prover -- not automatically, in the
general case. Said that once the proof-of-discord is
established by a peer, the others can autonomously validate the proof
and revert to a previous state if needed and according to the
semantics defined in case of contradictions (if any). In the following
we abstract the run-time management of Pluralize 
through the operational semantics of the language, leaving out of the
scope of this paper its actual implementation.

    

\subsection{Blockchain Specification} \label{ssec:adt}
  In~\cite{DBLP:journals/corr/abs-1802-09877} the authors present an
 abstract data type to formally model the blockchain object and its admissible
 behaviors. A blockchain is specified as a {\it BlockTree ADT}
 (BT-ADT)--specifying the blockchain object -- augmented with a {\it Token
 Oracle ADT} ($\Theta-$ADT) -- specifying the validation mechanism of blocks composing
 the blockchain. In such context, the validation mechanism is made explicit due to its importance, that is, it is the main actor in determining the 
 blockchain object behavior. That abstraction of the blockchain that
 explicitly formalized the validation mechanism enables us to
 formalized the enrichment of the validation mechanism to manage the
 autonomous and unreliable guards. Moreover, the ADT enables to consider
 the properties of the blockchain as a shared data structre and its
 interface as an execution models. The benefits is the guarantee to
 consider in a same formalization the consistency of the host
 blockchain and the formalization of its enrichment. As a consequence,
 Pluralize is technology agnostic as it can be parametrized by the
 host blockchain BT-ADT intanciation.
 
 \paragraph{The BlockTree ADT: BT-ADT.}
 In~\cite{DBLP:journals/corr/abs-1802-09877}, the data structure
 implemented by blockchain-like systems is a directed rooted tree
 called {\it BlockTree}. BT-ADT specifies a blockchain-like system
 through the specification of the semantic of the $read$ and $append$
 operations. BT-ADT is parametrized by (i) the predicate $P$ that
 has to be satisfied to validate a block and (ii) the selection
 function $f$ that selects, in a BlockTree $bt$ the blockchain
 $b_h$ where the next append has to be performed. 

\paragraph{Token oracle: $\Theta-$ADT.}
 The $\Theta-$ADT formalizes the interaction between the validation of
 blocks and their publication in the blockchain data structure.  
 {In few words, the Token oracle is a way to abstract the process of appending new blocks, proposed by agents, to the blockchain. Either only one block can be appended to the same block (already in the blockchain) or more than ones, in both cases all blocks has to be valid. The Oracle provides tokens to agents willing to append a new block but limits the number of tokens that can be consumed, i.e., a token is consumed when a new block is appended. In such a way the Oracle manages the number of blocks that can be appended to the same block already in the blockchain. More formally,} in~\cite{DBLP:journals/corr/abs-1802-09877}, the authors
 abstract this implementation-dependent process by assuming that an
 agent obtains the right to chain a new block $b_l$ to $b_h$ if it
 successfully gains a token $tkn_h$ from $\Theta-$ADT for such
 block. A token can be consumed at most once and the proposed block
 $b_l$ along with $tkn_h$, denoted $b_l^{tkn_h}$ is considered as
 valid. That means it can be appended to $b_h$. More formally,
 $\Theta-$ADT specifies two operations {\sf getToken} and {\sf
 consumeToken}. {\sf getToken} inputs two blocks $b_h$ and $b_l$ where
 $b_l$ is the block to append to $b_h$ which is the last of $f(bt)$.
 {\sf getToken} outputs $b_l^{tkn_h}$ which satisfies $P$, in other
 words, it is a valid block.  {\sf consumeToken } inputs a valid block
 $b_l^{ tkn_h}$ and enables the consumption of tokens. A maximum
 number of tokens $k$ for a block can be consumed \footnote{More into
 details and out of the scope of this paper,
 in~\cite{DBLP:journals/corr/abs-1802-09877} the authors defined two
 different typologies of $\Theta$-Oracle. The Prodigal Oracle,
 $\Theta_P$-Oracle in which an unbounded number of tokens can be
 consumed for each block, and the Frugal Oracle,
 $\Theta_{F,k}$-Oracle, that allows at most $k$ tokens to be consumed
 for each block, i.e., at most $k$ blocks can extend the same block.}.

 \paragraph{BT-ADT augmented with $\Theta-$ADT.}
 $\mathfrak{R}(BT-ADT, \Theta)$ BT-ADT augmented with $\Theta-$ADT is
 a $refinement$ that, once instantiated with a predicate $P$ and a selection function $f$, gives the
 specification of a blockchain-like system $S$ as an execution
 model. Specifically, in~\cite{DBLP:journals/corr/abs-1802-09877}, the
 authors define a refinement of the $append(b_l)$ operation of the BT-
 ADT with the oracle operations which triggers the {\tt
 getToken}$(b_h \leftarrow${\tt last\_block}$(f (bt)), b_l)$ operation
 as long as it returns a token on $b_h$ , i.e., $b_l^{tkn_h}$ which is
 a valid block. Once obtained, the token is consumed and the append
 terminates, i.e. the block $b_l^{tkn_h}$ is appended to the block
 $b_h$ in the blockchain $f(bt)$. We indicate such append with the
 following notation $f (bt)|_h^\frown {b_l }$ (slightly simplified
 with respect to the one in ~\cite{DBLP:journals/corr/abs-1802-09877})
 that we call {\it concrete append} in this paper. Notice that those
 two operations and the concatenation occur atomically.
}

        \section{ Pluralize smart contracts: language and execution model } \label{sec:mf}
		{ {\it Pluralize} is a formal framework for higher level blockchain transaction 
equipped with a smart contract formal language, called {\it
Plurality}.  In this section, we give a brief overview of Plurality,
just to illustrate the transfer of ownership between agents and the
two kind of guards. In this language, agents submit transactions to
the blockchain and are notified when transactions are published in the
blockchain. The notification mechanism is out of the scope of this
paper but introduce in order to illustrate the expressivness of
Plurality.  We also assume that Pluralize features a synchronisation
mechanism that enables to ensue causal order between transaction of a
same smart contract. At execution, we assume that the notification
mechanism ensures the synchronization that are work in progress for
Pluralize.

A transaction is a transfer of ownership of a quantity of
tokens \footnote{Please notice that this concept of token is different
from the one used by the Token Oracle.}, from an agent \textit{source}
to an agent \textit{sink} guarded by a property that has to be
validated.  A {\it Plurality} smart contract is a distributed protocol
of such transactions between agents in a network.  As a dialect of
Cyberlogic, {\it Plurality} is a shallow-embbeding formal language in
the Coq Theorem prover, its semantics is then given in terms of Coq
language.  However, its execution models abstracts (i) the host
blockchain protocol, (ii) validation mechanism and (iii) the
enrichment of {\it Pluralize}. As {\it Plurality} smart contract can
benefits the extraction mechanism of The Coq Theorem Prover, once
extracted in an executable language, a {\it Plurality} smart contract
will be executed in {\it Pluralize} execution framework: the host
blockchain equipped with an enriched validation mechanism to handle
claimed and closed guards validations.

 Section~\ref{sec:syn} presents the core of {\it
 Plurality}. Section~\ref{sec:ex} gives some examples of smart
 contracts. Section~\ref{sec:pmec} presents {\it Pluralize}
 execution model for {\it Plurality} smart contracts and details the
 operational semantics of transaction.

 \subsection{The Kernel of the {\it Plurality} Formal Language} \label{sec:syn}

 \paragraph{The syntax elements of the Kernel}
  The BNF of the kernel of {\it Plurality} is defined as follow: \\* 
  \begin{tabular}{lll}
   $agent$&$ \mathbb{A}::= $&$A,B, \Omega_i$\\
   $guard$&$gd$ \\
   $endorsement$&$claim ::=$&$\attest{\Omega_i}{gd} $ \\
   $transaction$&$ T::=$& $ \trans{A}{q}{gd}{B} | \trans{A}{q}{claim_{ik}}{B}$ \\
   $actions$&$ C::= $&$ \iss{\mbox{\small $x$}}{T} ~| \siss{\mbox {\small $\vec{x}$}}{\mbox{\small $x$}}{T} $ \\
   $smart contract $&$P::=$&$ C;\ldots;C $ \\
\end{tabular} \\

 {\it Plurality} is a dialect of Cyberlogic enabling each agent to
  submit transactions to the blockchain. Agents of {\it Plurality} are
  Cyberlogic authorities. The syntax distinguishes two kind of agents in
  order to explicit the accountability of both: agents involved in
  a transaction (as source or sink) and external agents, oracles
  denoted $\Omega_i$, that endorse guards. Transactions are guarded
  transfers of ownership from the wallet of $A$ to the wallet of $B$
  of an amount $q$ of tokens. The kernel of {\it Plurality} manages a
  unique kind of tokens, $\tau$.

\paragraph{Transaction and their guards.}
The syntax of {\it Plurality} distinguishes two kind of transactions
according to the type of guard. $\trans{A}{q}{gd}{B}$ is the
transfer of an amount of $q~\tau$ from $A$ to $B$ guarded by the closed guard
$gd$. $\trans{A}{q}{claim}{B}$ is the transfer of an amount of $q~\tau$ from
$A$ to $B$ guarded by the claimed guard $claim$.

  \paragraph{Submission of transactions.}  An agent that owns a wallet
 in the blockchain can submit transactions in two different
 manners. $\iss{{\small x}}{T} $ is the submission of the transaction
 $T$ by the source of $T$, the transaction is bound to the identifier
 $x$. We assume that each identifier is unique.
 $\siss{{\small \vec{x}}}{{\small x}}{T}$ is the submission of $T$ by
 its source after that all transactions of $\vec{x}$ have been
 published in the blockchain.

 We denote as $|W|$ the current balance of the agent $W$. For example,
 $\iss{{\small x}}{\trans{A}{5}{|B|<2~}{B}}$, is the submission by the
 agent $A$ of the transaction $x$ of $5~\tau$ from $A$ to $B$ guarded
 by the closed guard $|B|<2$. This transaction can either be validated
 or not. If it is validated, $A$ can publish it on the blockchain.

 $\iss{{\small x}}{\trans{A}{5}{\attest{\Omega_{Bk}}{account(B)<2~}}{B}}$, 
 is the
 submission by the agent $A$ of the transaction $x$ of $5~\tau$ from
 $A$ to $B$ guarded by the claimed guard
 $\attest{\Omega_{Bk}}{account(B)<2~}$, the bank $\Omega_{Bk}$
 endorses that $B$ has less than $2~\$$ on is bank account. This
 transaction can either be validated or not. If it is validated, $A$
 can publish it on the blockchain. Here, $\Omega_{Bk}$ is an oracle
 that endorses the guard of the transaction.
 
 $\siss{{\small [x]}}{{\small
  y}}{\trans{B}{3}{\attest{K_t}{bill\_day}~}{C}}$, is the submission
  by $B$ of the transaction $y$ of $3~\tau$ from $B$ to $C$ if the
  time oracle $K_t$ states that it is the $bill\_day$. To submit $y$,
  $B$ has to wait that $x$ has been published. We denoted $[]$ as a list and $[x]$ 
  the list that contains only the element $x$.

 The validation mechanism and the publication of 
 transactions are detailed in the section~\ref{sec:pmec}.

 \subsection{{\it Plurality} Smart Contract examples} \label{sec:ex}
This section aims at presenting the different kind of guards in the
core of {\it Plurality} concentrated on guards and their
validations. Please notice that the transactional expressivness of
{\it Plurality} is very light and is a work in a progress out of the
scope of the paper. In particular {\it Plurality} does not yet feature
alternative choices. Moreover, we do not have detail temporality of
guards for the sake of clarity, but as a dialect of Cyberlogic, {\it
Plurality} and more specifically, the claimed guards are ables to
reason about time thanks to the specific $K_t$ authority of
Cyberlogic: the time oracle. Finally, as detail bellow, the semantics
of {\it Plurality} is given in CPS style, this formalism enables to
disjonct concurrent execution. According the the consistency of the
host blockchain, a {\it Plurality} smart contract will be able or not
to present divergent history. for the sake of simplicity, in the following example,
we assume that $\mathbb{B}$, the host blockchain is an ideal blockchain.

 \paragraph{The Fair Pocket Money.}
A family $F$ uses the blockchain to distribute the pocket
money of their two kids $A$ and $B$ in $\tau$. Their pocket money
budget, $50~\tau$ is stored in the wallet $W$. The distribution of the
pocket money is fair: both receive $20~\tau$. \\*
\begin{tabular}{l}
  $ \iss{x}{\trans{F}{50}{\top}{W}};~\siss{x}{y}{\trans{W}{20}{\top}{A}};~\siss{x}{z}{\trans{W}{20}{\top}{B}};$ \\
\end{tabular}

Note that in this smart contract guards are always at true, but a
specific order on transactions is specified. The smart contract
contains 3 transaction submissions: $x$ is the submission of the
transfer of the pocket money budget from the family $F$ to its pocket
money wallet $W$, $y$ is the submission of the transaction of the part
of the pocket money to $A$ after the publication of $x$ and $z$ is the
submission of the transaction of the pocket money to $A$
after $x$ publication.

 \paragraph{The Studious Pocket Money.}  The family $G$ distributes
 the pocket money according to school merit: if the grate of the kid
 at school is greater to $10$ then the kid receives $20$ otherwise
 only $10$. In this example, we suppose that the current balance of
 $S_A$, $|S_A|$ contains a amount of $\tau$ that applies to
 $as\_grate$ encodes the grate of $A$. The transaction $s$ is the
 submission from the school to $S_A$ that updates the $A$ grate.

 \begin{tabular}{l}
 $\iss{x}{\trans{F}{50}{\top}{W}} ; $\\
 $\siss{[x;s]}{y}{\trans{W}{20}{ as\_grate( |S_A|) >10}{A}}  $ \\
 $\siss{[x;s]}{z}{\trans{W}{10}{ as\_grate( |S_A|) =<10 }{A}} $ \\
 \end{tabular}\\

 As for the fair pocket money smart contract, in $x$ the family
 provides $50$ in their pocket money wallet. $y$ is the submission of
 the transaction from $W$ to $A$ $20~\tau$ under the condition that
 her rate is superior to $10$. The submission of $y$ by $W$ is
 possible only if the transaction $x$ and $s$ has been
 published. Hence, if the guard $ as\_grate( |S_A|) >10$, which is a
 closed guard is validated, $W$ publishes $y$: the wallet of $A$ is
 updated of $20$ more $\tau$.  $z$ very similar to $y$ except that its
 guard is $ as\_grate( |S_A|) <=10$ and in case of validation followed
 by publication: the wallet of $A$ is only updated of $10$ more
 $\tau$.

 \paragraph{The evidential Pocket Money.}
 The family $L$ will only give the pocket money to a kid if it
 obtains its driving license.
 
\begin{tabular}{l}
  $\iss{x}{\trans{F}{50}{\top}{W}~as~T_x} ; $\\
  $\siss{[x]}{a}{\trans{W}{20}{\attest{\Omega_{X}}{license(A)}}{A}}  $ \\
\end{tabular} \\
  
  $a$ can be submitted only if $x$ has been published. The validation
of $a$ is the validation of $\attest{\Omega_X}{license(A)}$, the
endorsment of the agent $\Omega_X$, an oracle, that $A$ has her
license. Suppose that none of the already published transactions can be
contradicted by $\attest{\Omega}{license(A)}$. Hence, $a$ can be
validated by the validation mechanism and published by $A$. 

 \paragraph{The Competitive Pocket Money.}  The family $M$ has 2
 kids, $A$ and $B$, in the same class $C$.  $M$ gives a pocket money
 to a kid only if he is the top of the class.  In the first version,
 the rank of a kid in $C$ is computable in the blockchain by the
 predicate $rank(C,X)$ the rank of the student $X$ in
 $C$. \footnote{That is feasible if for any student of $C$, $X$, her/his
 marks are readable and used to compute the ranking  of students of $C$, this
 computation is closed to the blockchain}. $rank$ ensures that each
 position in the rank owns to only one student.

 \begin{tabular}{l}
  $\iss{x}{\trans{F}{50}{\top}{W}} ; $\\
  $\siss{[x]}{a}{\trans{W}{20}{rank(C,A)=1}{A}}  $ \\
  $\siss{[x]}{b}{\trans{W}{20}{rank(C,B)=1}{B}}  $ \\
\end{tabular} \\
 
 In this smart contract $rank(C,X)=1$ is decidable and only one $X$
 can satisfy $rank(C,X)=1$. Hence, if $x$ is published, it is not
 possible that both, $a$ and $b$ are validated: either $A$ or $B$ does
 not receive a pocket money -- rank is not $1$ for both $A$ and $B$ --
 or only one of them receives it -- $A$ or $B$ rank is $1$.
 
 In the second version, $\Omega_{s}$ endorses the kids rank. \\
 \begin{tabular}{l}
  $\iss{x}{\trans{F}{50}{\top}{W}~} ; $\\
  $\siss{[x]}{a}{\trans{W}{20}{\attest{\Omega_s}{rank(C,A)=1}}{A}}  $ \\
  $\siss{[x]}{b}{\trans{W}{20}{\attest{\Omega_s}{rank(C,B)=1}}{B}}  $ \\
\end{tabular} \\
 
 Suppose that $x$ and $a$ have been published. $b$ can also be
 submitted.  Suppose now, that $\Omega_{s}$ endorses that $B$ is the
 first in the rank : $\attest{\Omega_s}{rank(B)=1}$. If $b$ is
 validated, it would contradict $a$: either $A$ or $B$ can be ranked
 as the first student in the class but not both of them.  In that
 case, the validation mechanism has to prevent the publication of $b$.
 The validation will formally detect that $b$ contradicts $a$ and $b$
 will not be validated. This detection is computed by the
 proof-of-discord, the part of the validation mechanism
 dedicated to the validation of claimed guards. Notice, that this
 detection does not indicate which between $a$ or $b$ is false. In
 case of a proof-of-discord, the conflict has to be
 managed. The proof-of-discord is detailed in
 Section~\ref{sec:pmec}.

 \subsection{{\it Pluralize} Formal Execution Models} \label{sec:pmec}
 
Semantically, a {\it Plurality} smart contract executes on the
execution model of {\it Pluralize}. {\it Pluralize} is a combination
of its host blockchain $\mathbb{B} $ and the validation mechanism
$\mathbb{V}$ that validates guards of submitted transactions. {\it
Plurality} mainly manipulate transactions, then the crucial point of
the operational semantics of {\it Plurality} is the specification of
the different steps of a transaction execution. We call these steps
the {\it execution steps}: (i) the submission of a transaction by its
source, (ii) the validation of a submitted transaction by
$\Theta_{\mathbb{B}}$, and (iii) the publication a validated
transactions in the blockchain by its source. While each step can be
performed by different agents, their executions must happen in a
sequential specific order. A relevant technique to specify such
operational semantics is the {\it Continuation Passing Style},
CPS~\cite{Plotkin}.

 As a dialect of Cyberlogic which is implemented in the Coq theroem
Prover, {\it Plurality} is a functional language -- a programing
language with functions as first-class citizen based on
$\lambda-calculus$~\cite{lcalc}.  A relevant technique to specify concurency in
functional languages is the {\it Continuation Passing Style},
CPS~\cite{Plotkin}. CPS explicits the evaluation order, this is truly
relevant for actions that have to be executed in a specific order but
that are performed by different agents. We specify each execution step
as a Cyberlogic continuation that manipulate the execution model.

A continuation captures the context at some given point in the
program: it implicitly records the current instruction and the local
state. Continuations are therefore perfectly suited to implementing
concurrency. If the call to function were a cooperation point in a
threaded program for instance, saving its continuation would be enough
to resume execution after a context switch.  Continuations are most
often used in functional programming languages. Some of them, like
Scheme~\cite{Scheme} or Scala~\cite{Scala}, provide first-class continuations
with control operators, such as {\tt call/cc} or {\tt shift} and
{\tt reset} respectively, that allow a program to capture and resume its own
continuations. Cooperative threads and other concurrency constructs
are then built on top of these operators~\cite{ScalaThread}.  In
functional languages that do not provide first-class continuations,
continuations are encoded using other features such as first-class
functions or monads. These constructs can then be used to implement
concurrency libraries: concurrency monads in Haskell~\cite{HaskelConcurentMonad}, or
lightweight lwt threads in OCaml~\cite{VouillonLwt}. In Javascript, callback are
continuation and in new version Promise and Async/Await are some kind of
explicit concurrent monad.

 The semantics of {\it Plurality} is given in CPS-style Cyberlogic
formula and is built upon Coq terms semantics (as it is a
shallow-embbeding implementation). In other words, a {\it Plurality}
smart contracts is correct if the type checker of Coq checks it: that
brings a high level of formal guarantee and gives {\it Plurality} a
well-founded semantics. The crucial point is the abstraction of the
execution models. Thanks to the blockchain ADT, that formalization is
quite elegant.

Further in this section we present {\it Pluralize} as an execution model
for {\it Plurality} transactions. First, we define the blockchain
specification of {\it Pluralize}. Then, we give the specification of
the execution steps in the execution model. Finally, we detail the
validation mechanism $\mathbb{V}$.

  \paragraph{{\it Pluralize} blockchain specification.} The
  specification of {\it Pluralize} is the
  BT-ADT~\cite{DBLP:journals/corr/abs-1802-09877} instantiated with
  $f$ the selection function of $\mathbb{B}$ and $\mathbb{V}$ the
  validation predicate. $\mathbb{V}$ validates both the closed guards
  thanks to the validation predicate of $\mathbb{B}$, and the claimed
  guards using proof-of-discord.  $\Theta_{\mathbb{B}}$ is the refinement
  of BT-ADT$(f,\mathbb{V})$ with $\Theta$ the Token oracle ADT
  $\mathfrak{R}($ BT-ADT$(f,\mathbb{V}),\Theta)$.

 \paragraph{Transfer of ownership as a Cyberlogic formula.} 

 We define $Account$ as the interpretation of a {\it Plurality}
 transaction $T$. $Account$ defines the semantics of {\it Plurality}
 transaction in Cyberlogic. As Cyberlogic is a shallow embeding in
 Coq, the semantics of {\it Plurality} is defined in the semantics of
 CiC~\cite{Cic}. Hence,a {\it Plurality} transaction becomes: \\*
 $Account(\trans{A}{q}{c}{B})= \attest{\Theta}{\mathbb{V}(c)=\top} \rightarrow \attest{A}{updates(A,q,B)}$
 where $updates(A,q,B)$ is the transfer of ownership of $q~\tau$ from
 the wallet of $A$ to the wallet of $B$.

  \paragraph{Submission continuation.} $\kappa_{sub}$ is the
  continuation that defines the semantics of the submission of a
  transaction.  \begin{tabular}{l}
  $\kappa_{sub}(T~,~\kappa_{\mathbb{V}} ) =${\sf ~let }$~b_l =
  Account(T)${\sf ~in }$~ \kappa_{\mathbb{V}}
  (b_l)$ 
  \\ \end{tabular} 
  \\ $\kappa_{sub}$ inputs a transaction $T$
  and a validation continuation $\kappa_{\mathbb{V}}$.  $\kappa_{sub}$
  makes a logical formula $b_l$ from the transaction $T$ thanks to the
  function $Account$, and passes $b_l$ to $\kappa_{\mathbb{V}}$.
  Hence, the semantics of a submission of the transaction $T$ is the
  application of $\kappa_{sub} (T)$ that waits for the application to
  a validation continuation $\kappa_{\mathbb{V}}$ that can be
  processed by $\Theta_{\mathbb{B}}$.

 \paragraph{Validation continuation.} $\kappa_{\mathbb{V}}$ is
the continuation of validation that inputs  the
Cyberlogic formula of a submitted transaction $b_l$ and a continuation of
publication $\kappa_{pub}$. $\kappa_{\mathbb{V}}$ requests a token
$b_l^{tkn_k}$ to $\Theta$. If the token is obtained,
$\kappa_{\mathbb{V}}$ applies it to $\kappa_{pub}$. \\*
 \begin{tabular}{l}
   $\kappa_{\mathbb{V}} (b_l~,~\kappa_{pub}) =  $ \\
  {\sf {\small case getToken}}$(b_l,${\sf {\small ~last\_block}}$(f(bf)))= b_l^{tkn_k}:~\kappa_{pub} (b_l^{tkn_k});$ \\
 \end{tabular}

 \paragraph{Publication continuation.} $\kappa_{pub}$, the publication
  continuation is a consumption of a token of $\Theta$ and the
  concrete append of the validated transaction in the
  blockchain.

  \begin{tabular}{l} $\kappa_{pub} (b_l^{tkn_k}) = $ {\sf
  consumeToken}$(b_l^{tkn_k});  f (bt)|_k^\frown {b_l }.$ \\
\end{tabular}
  
  \paragraph{$\mathbb{V}$ and the proof-of-discord.}
     We detail the specification of
     $\mathbb{V}$, the validation mechanism of both kind of
     guards as follows: \\
 \begin{tabular}{l}
       $\mathbb{V}(b_l,b_h, \kappa_{C}) = $ \\
       \begin{tabular}{ll}
        ~~~&$P_{app}(b_l) \&$ \\
        ~~~&$case~guard(b_l)=gd: P_{gd}(gd)$ \\
        ~~~&$case~guard(b_l)=claim_{ik}:~\&~pod(b_l,b_h)= (\top,\varnothing):~\top$ \\
        ~~~&$case~guard(b_l)=claim_{ik}:~\&~pod(b_l,b_h)= (\bot,\mathbb{C}_{ik}):~\kappa_{C} (\mathbb{C}_{ik})$  \\
       \end{tabular}\\
      $poa(b_l~where~guard=claim_{ik},b_h,\kappa_{C}) = $ \\
       \begin{tabular}{ll}
        ~~~&{\small{\tt ~let }}$~\Gamma:=${\small{\tt ~make\_context}}$(b_h);$ \\
        ~~~&{\small{\tt ~case }}$~\Pi(\Gamma \vdash \neg claim_{ik})= \varnothing: (\top,\varnothing);$ \\
        ~~~&{\small{\tt ~case }}$~\Pi(\Gamma \vdash \neg claim_{ik})= \mathbb{C}_{ik}: (\bot,\mathbb{C}_{ik}) ;$ \\
       \end{tabular}
     \end{tabular} \\

 Consider $P_{\mathbb{B}}$ as the validation predicate of $\mathbb{B}$.
 $P_{\mathbb{B}}$ is the conjonction of two predicates: $P_{app}$, the
 predicate that ensures the validation of the append conditions that
 are specific to $\mathbb{B}$ and $P_{gd}$, the validation of closed guards.

 If the guard is an closed guard $gd$, then $\mathbb{V}$ verifies if
 $P_{\mathbb{B}}(gd)= \top$.  In case of an claimed guard $claim_{ik}$,
 $\mathbb{V}$ verifies if $P_{app}$ is satisfied by the submitted
 transaction and tries to  obtain a proof-of-discord for
 $claim_{ik}$ from {\small{\tt ~pod}}. In order to manage proof-of-discord, 
 the validation mechanism inputs a  $\kappa_{C}$, the manager of
 conflict, which is unspecified in this paper. {\small{\tt pod}} inputs: (i)
 $b_l$, the Cyberlogic formula of a submitted transaction and (ii) $b_h$,
 the last block of $f(bt)$. {\small{\tt ~pod}} makes a proof
 context from $b_h$ and asks a formal certificate for
 $\Gamma \rightarrow \neg claim_{ik}$ \footnote{The verification that
 $claim_{ik}$ is true in the evaluation context $\Gamma$} to $\Pi$, a
 specific agent that makes a proof in Coq. If $\Pi$
 fails then $claim_{ik}$ is valid, if $\Pi$ provides a formal
 certificate: $\mathbb{C}_{ik}$ then $claim_{ik} $ is unvalid and
 {\small{\tt ~poa}} returns the certificate.
}

	\section{Related Work} \label{sec:relatedwork}
		{ 
  In~\cite{cyber18}, the authors higlighted the need of accountability
  for smart contracts that act as evidential protocols and proposed to
  use the authority algebra of Cyberlogic~\cite{RS:HCSS03}, but
  without defining a formal language supporting it.

  Scilla~\cite{DBLP:journals/corr/abs-1801-00687} is a formal
  intermediate language for Ethereum smart contracts that is a
  shallow-embedding in Coq. As {\it Plurality}, the Scilla semantics
  is defined in the semantics of Coq in a CPS style and Scilla smart
  contract are certified-by-designed.  As Solidity smart contracts,
  Scilla smart contracts are stateful executable object-based code
  hosted on the blockchain that behaves as an autonomous server which
  has to be invoked by clients. Contrarily to Scilla, {\it Pluralize}
  is suitable to digitalize real-life transactional protocols: it
  focuses on smart contracts that are guarded transfers of ownership
  between agents with accountability as first class
  citizens.

Typecoin~\cite{Crary:2015:PAC:2813885.2737997} is a protocol built
upon Bitcoin, that transfers ownership of properties instead of
ownership of amount of token: Typecoin implements proof-carrying
authorization upon the Bitcoin protocols. That means that it enables
to verified the authorization -- script that involves cryptographic
signature verification. As in {\it Pluralize}, Typecoin sees the
transaction as a transfer between agents, but reasoning on
authorization instead of accountability as we do in {\it
Pluralize}. Proof-carrying authorization~\cite{Appel1999} is more
dedicated to permission and access properties than accountability and
reliability.  Typecoin used opcodes and the existing validation
mechanism of bitcoin, even if it offers formal guarantee it suffers of
the same lack of expressivness than bitcoin script. {\it Pluralize} by
enriching the validation mechanism of the host blockchain brings more
formal guarantees and increase the reliability of smart contracts.

 BitML~\cite{BitML} is a domain-specific language for smart contracts
 compiled to Bitcoin script. It is defined as a process calculus that
 is sufficiently expressive to cover most applications proposed in the
 literature for Bitcoin script. The DSL comes with a symbolic
 semantics. The paper also defines a computational model and
 establishes a relationship between the symbolic semantics and the
 computational model, which they call coherence. Based on this notion
 of {\it coherence}, the paper establishes the soundness of the
 compiler that takes the DSL into Bitcoin script. BitML offers formal
 guarantees and higher level expressiveness to implement current
 Bitcoin smart contract.  {\it Pluralize} tackles smart legal contract
 at a higher-level by enriching the validation mechanism to handle
 claimed and closed guards that are delegated to Oracle in BitML as
 in the current Bitcoin smart contract. Thanks to the blockchain ADT,
 the coherence in {\it Pluralize} is inherits from the host blockchain
 and there is no need to establish an equivalence between the symbolic
 and execution model, as the ADT enables to formalize both in a same
 abstraction. That enables {\it Plurality} to target different
 blockchain protocols and makes it be technology agnostic.  On the
 other hand, BitML explicits in its language and semantics the
 concurrency of its protocols while, in its current version, {\it
 Plurality} does not offer explicit concurrency choices in its
 syntax. However, using CPS to define its semantics eases this kind of
 improvement: CPS already allows to express concurrent
 semantics. Naturally, as {\it Plurality} aims at provide a
 higher-level smart legal contract language, it also have to feature
 concurrency in its syntax.

}

	\section{Conclusion} \label{sec:conclusion}
		In this paper we presented {\it Pluralize}, a formal framework for
accountability of blockchain transactions that extended Bitcoin
paradigm to guards whose evaluation is not closed to the blockchain.
{\it Pluralize} features a formal language, enabling the
implementation of trustworthy smart contracts, called {\it
Plurality}. {\it Pluralize} enables formal features in order to gain
trust in smart contracts from their specification to their execution
on the blockchain. {\it Plurality} is a dialect of Cyberlogic
benefiting from the Cyberlogic theorem prover: its smart contracts are
certified-by-design and their properties can be formally verified. The
idea behind Pluralize is to concialate the bitcoin-like smart contract
with the smart legal contracts as define by~\cite{SCdef} and that inspires
the industrial predictive usage.
Finally, {\it Pluralize} features proof-of-discord thanks to
the possibility to trace back the chain of trust by interpreting
transactions as Cyberlogic claims and either certificate that the
blockchain is right or provide the potential guilty claims. Note that
this work has been motivated by the needs identified during the
development of an industrial prototype~\cite{gurcan:cea-01807039}.

As future work we envisage to generate dynamic monitors for
claimed guards. Monitors will be embedded in the run-time
environment of the language for run-time identification of
conflicts. Note that thanks to our Coq-enabled framework run-time
monitors can be generated as correct-by-construction and a certification of their
computation can be provided as a locally-checkable proof. This turns
out to obtain proofs about transaction validation/invalidation
autonomously checkable by any node in the network.

 Another important future work is to increase the expressivity of {\it
 Plurality} for ownership transfers.  First, transfer have to be
 generalized to heterogeneous tokens thanks to a token algebra.
 Second, the validation mechanism has to handle the validation of {\it
 smart contract computational} guards.  Finally, for their deployment,
 these enriched smart contracts might require side-chains or
 multi-chain components for complex validation processes and to store
 heterogeneous tokens. Hence, Pluralize executions model should be
 extended to several blockchains.  Correctness of such extension will
 depends on the consistency of blockchains and consistency of their
 interaction. Formalization of those blockchains and their interaction
 requires to be able to compose BT-ADT instantiations.





\bibliographystyle{abbrv}
\bibliography{biblio} 
\newpage
\appendix 
\end{document}